\documentclass[12pt,twoside,a4paper]{article}

\usepackage{graphicx}
\usepackage{bigstrut}
\usepackage[margin=1in]{geometry}
\usepackage{fancyhdr}
\usepackage[dvipsnames]{xcolor}
\usepackage{mathtools}
\usepackage[T1]{fontenc}
\usepackage{sourcesanspro}
\usepackage{titlesec}
\usepackage{bbm}
\usepackage{bigstrut}
\usepackage{multirow}
\usepackage{ctable}
\usepackage{colortbl}
\usepackage{enumerate}
\usepackage{tabularx}
\usepackage{booktabs}
\usepackage[colorlinks,allcolors=blue]{hyperref}
\usepackage[utopia,cal=cmcal]{mathdesign}
\usepackage{erewhon}
\usepackage[font={footnotesize,sf},labelfont=bf]{caption}
\usepackage{titling}

\newcommand{\orcid}[1]{\href{https://orcid.org/#1}{\includegraphics[height=\fontcharht\font`M]{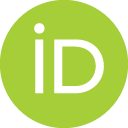}}}

\newtheorem{theorem}{Theorem}[section]
\newtheorem{corollary}{Corollary}

\newcolumntype{Z}{>{\centering\arraybackslash}X} 

\titleformat{\section}{\sffamily\Large\bfseries}{\thesection}{1em}{}
\titleformat{\subsection}{\sffamily\large\bfseries}{\thesubsection}{1em}{}

\pretitle{\begin{flushleft}\sffamily}
\posttitle{\end{flushleft}\sffamily}
\preauthor{\begin{flushleft}\sffamily}
\postauthor{\end{flushleft}\sffamily}
\predate{}
\postdate{}

\thanksmarkseries{fnsymbol}

\title{\LARGE\sffamily\bfseries{Deconstructing the role of myosin contractility in force fluctuations within focal adhesions}}
\author{\sffamily\bfseries Debsuvra Ghosh\thanks{\sffamily Department of Physical Sciences, Indian Institute of Science Education and Research Mohali, Sector 81, Knowledge City, S. A. S. Nagar, Manauli 140306, India} \,\orcid{0000-0003-2471-203X} ,\, Subhadip Ghosh\thanks{\sffamily Department of Physics, Faculty of Science, University of Zagreb, Bijenička cesta 32, 10000 Zagreb, Croatia},\, Abhishek Chaudhuri\thanksmark{1} \,\orcid{0000-0003-0458-9601} \,\thanks{\sffamily Corresponding author. Email: \href{mailto:abhishek@iisermohali.ac.in}{abhishek@iisermohali.ac.in}}}

\date{}

\begin{document}
\maketitle

\begin{abstract}
  \sffamily
  Force fluctuations exhibited in focal adhesions (FAs) that connect a cell to its extracellular environment, point to the complex role of the underlying machinery that controls cell migration. To elucidate the explicit role of myosin motors in the temporal traction force oscillations, we vary the contractility of these motors in a dynamical model based on the molecular clutch hypothesis. As the contractility is lowered, effected both by changing the motor velocity and the rate of attachment/detachment, we show analytically in an experimentally relevant parameter space that the system goes from decaying oscillations to stable limit cycle oscillations through a supercritical Hopf bifurcation. As a function of motor activity and the number of clutches, the system exhibits a wide array of dynamical states. We corroborate our analytical results with stochastic simulations of the motor-clutch system. We obtain limit cycle oscillations in the parameter regime as predicted by our model. The frequency range of oscillations in the average clutch and motor deformation compares well with experimental results.
\end{abstract}

\pagestyle{fancy}

\fancypagestyle{firstpage}{
    \fancyhf{}
    \fancyhead[L]{\sffamily arXiv PREPRINT \, $\vert$ \, {\color{BrickRed}RESEARCH ARTICLE}}
    \fancyfoot[L]{\sffamily\scriptsize Debsuvra Ghosh et al., arXiv:2011.13767v2 \date}
    \fancyfoot[R]{\sffamily \bfseries \thepage}
}

\thispagestyle{firstpage}

\fancyhf{}
\fancyhead[L]{\sffamily arXiv PREPRINT \, $\vert$ \, {\color{BrickRed}RESEARCH ARTICLE}}


\fancyfoot[L]{\sffamily\scriptsize Debsuvra Ghosh et al., arXiv:2011.13767v2 \date}
\fancyfoot[R]{\sffamily \bfseries \thepage}

\section*{Introduction}
Cellular migration plays a critical role in a host of biological processes starting from embryonic development to the immunological response of the cell as well as wound healing~\cite{petrie2009random,de2005integrin,paszek2005tensional,yamaguchi2005cell,schwarz2013physics,paluch2016focal}. The disruption of cellular migration can lead to cancer metastasis and other chronic inflammatory diseases. The process of cell migration involves the sophisticated regulation of the machinery of the actomyosin complex comprising of the actin filaments and the myosin motors, the adaptor proteins which are linked to the actin and their subsequent linking to the transmembrane proteins which connect to the cell microenvironment~\cite{case2015integration,elosegui2018control,plotnikov2013guiding, blanchoin2014actin,charras2014physical}. Both in-vitro and in-vivo experiments have provided in-depth understanding of the role of each of the individual components of this extensive machinery as well as how they work in unison. In-vitro studies on two-dimensional substrates have provided valuable information about how cells interact with the substrate and move on it and how cell speeds are modulated depending on several mechanical and chemical cues. Although the various components of this process are well known, the measurement of the mechanical forces shows significant variability at the cellular level~\cite{kurzawa2017dissipation,meili2010myosin,rape2011regulation,plotnikov2012force}, making it imperative to decipher the key parameters that regulate them.

\begin{figure*}
	\includegraphics[width=\textwidth]{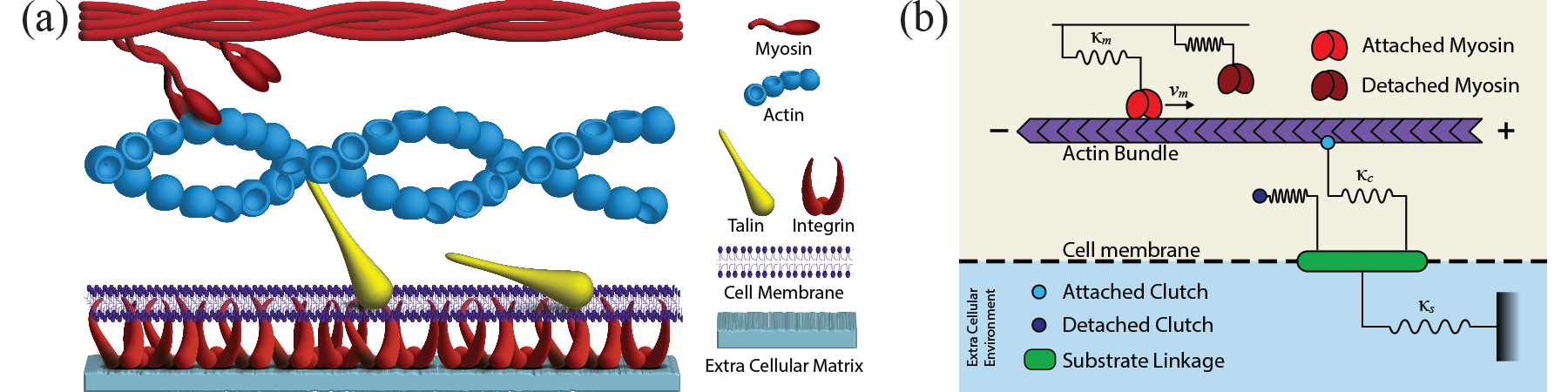}
	\caption{(a) Schematic of the cell migration machinery showing the myosin motors and actin bundle which constitute the cytoskeletal network. Adaptor proteins (talin) and transmembrane proteins (integrin) form the focal adhesions (FA) linking the cytoskeletal network to the extracellular matrix/substrate. (b) Motor clutch model showing the motors and clutches as elastic springs and an inextensible actin bundle with $\pm$ shows the anterograde/retrograde directions.}
	\label{fig:diagram}
\end{figure*}

The entire molecular assembly involved in cell migration, called the adhesion complex, is highly dynamic with a constant attachment/detachment kinetics between the various elements linking the cytoskeleton to the extracellular matrix. The combined effect of the myosin motors exerting contractile forces and the polymerisation of filamentous actin pushing against the cell membrane drives a `retrograde flow' of actin toward the centre of the cell. The `molecular clutch' hypothesis posits the focal adhesions as mechanical clutches that act as dynamic linkages in the transmission of forces between the actin filament and the transmembrane proteins, converting the retrograde flow to forward movement of the cell~\cite{elosegui2016mechanical} (Fig.~\ref{fig:diagram}(a)).

Apart from the in-vitro and in-vivo experimental studies, theoretical models have been proposed, which have proved helpful in understanding cell migration both at the cellular and molecular scales. Earlier theoretical studies~\cite{dimilla1991mathematical,macdonald2008kinetic,sabass2010modeling,schwarz2012united,chan2008traction,bangasser2013determinants,bangasser2013master,danuser2013mathematical,sens2013rigidity,bangasser2017shifting,craig2015model,leoni2017model,bressloff2020stochastic} have predominantly looked at the response of the cell to varying substrate rigidity and predicted a biphasic relationship between rigidity and force, i.e., forces first increase and then decrease with rigidity.  DiMilia et al.~\cite{dimilla1991mathematical} combined a visco-elastic-solid model for a cell and adhesion receptor-ligand binding kinetics for the adhesion bonds to predict how cell movement on a rigid substrate can vary with contractility and receptor-ligand kinetics. In certain parameter regimes, the cell speed was shown to be biphasic, with the maxima decided by a balance between contractility and adhesiveness. Using a force-based dynamic approach, Zaman et al.~\cite{zaman2005computational} developed a computational model for cell migration in 3-dimensional matrices. Similar to the situation in 2-dimensional substrates, a biphasic behaviour of cell speed with varying adhesivity is predicted. Transmembrane proteins such as integrins have been modelled as Hookean springs with detachment rates increasing with the load force, demonstrating that the clustering of proteins increases with the increase of the stiffness of the substrate. A stochastic motor-clutch model introduced by Chan and Odde~\cite{chan2008traction} was able to describe the load-and-fail characteristic of cellular force transmission, which has been observed experimentally in migrating cells. This model takes the force-velocity relationship of the myosin motors into account, incorporates the load and fail dynamics of cellular adhesions, and predicts an optimal stiffness of the substrate when the force transmission is maximal, and the actin retrograde flow is minimal.

Recent experiments using time-lapse traction force microscopy have shown that the local forces exerted by individual focal adhesions vary spatiotemporally, suggesting repeated tugging of the extracellular matrix/substrate~\cite{plotnikov2012force,plotnikov2013guiding}. Mature focal adhesions exist in two states: a stable state with spatially and temporally invariant traction and a dynamic state in which they fluctuate indicative of a tugging mechanism on the extracellular membrane. These force fluctuations are a possible molecular mechanism for the cell to tightly control cellular movement based on any environmental cues~\cite{wu2017two,gardel2010mechanical}. The physical understanding of the fluctuations within an integrated cell migration model is an open problem. Myosin contractility is one of the ideal candidates to give rise to these fluctuations, since earlier mathematical models have predicted that the collective activity of motors on elastic materials can lead to spontaneous oscillations in the activity of local contractile units~\cite{grill2005theory}.

In this paper, we theoretically explore the specific role of myosin II activity in the molecular clutch setting~\cite{meili2010myosin,escribano2014discrete,labouesse2015cell,thomas2015non,barnhart2015balance,barnhart2017adhesion,kobb2017tension,greenberg2016perspective,jung2019cell,recho2013contraction,wilson2010myosin}. Activity in myosin II is incorporated by both its attachment-detachment dynamics with the actin filament and the velocity of the attached myosin motor proteins (MPs). How does the variation of the activity of these motors affect the dynamical stability of the molecular clutch system? Further, can we quantitatively estimate the local fluctuations resulting from the spontaneous oscillations of these local contractile units? Unlike earlier theoretical models, we instead focus on the stability of the MP-actin-clutch sector only and show that a variation of the activity of myosin motors gives rise to a multitude of dynamical states. Specifically, for a wide range of experimentally accessible parameter space, the system exhibits spontaneous decaying oscillations in a stable spiral region crossing into a stabilised, oscillatory region via a supercritical Hopf bifurcation~\cite{lavi2016deterministic}. We also discuss the specific nature of these oscillations and their connection to the traction force fluctuations observed in the experiments. We incorporate stochasticity into the problem and show that the primary features of the model are retained.

\section*{Methods}
\subsection*{Model description}
We consider a geometric arrangement consisting of a filamentous actin bundle in the vicinity of myosin II motors and molecular clutches~\cite{chan2008traction,bangasser2013determinants,bangasser2013master}. The myosin motors are rigidly fixed at one end, while the other end attaches to the F-actin bundle and induces a retrograde flow by applying a force on the bundle. Molecular clutches have one end irreversibly attached to a substrate while the other end engages reversibly with the F-actin bundle and resist the retrograde flow. The force built up in the attached molecular clutches leads to a traction force that is balanced by the tension and deformation in the substrate as depicted in Fig.~\ref{fig:diagram}(b). In our model, we consider the motor-clutch and substrate sector as separate blocks with the substrate deformation solely governed by the dynamical force balance in the motor-clutch sector.

Myosin contractility and the resultant force generation is dependent on its attachment-detachment dynamics~\cite{stam2015isoforms,koenderink2018architecture}. Myosin motors are modelled as stretchable springs, which, due to energy consumption via the hydrolysis of ATP, undergo attachment-detachment dynamics to/from the F-actin bundle. One end of the spring is fixed while the other end attaches(detaches) to(from) the F-actin with rates $\omega_a$ and $\omega_d$, respectively. Following well-established theoretical approaches for molecular motors, their detachment rates are considered to increase exponentially with a load force $|f_l|$ as $\omega_d = \omega_d^0\exp{(|f_l|/f_d)}$, where $f_d$ sets the force scale and $\omega_d^0$ is the bare detachment rate. Extension $y^i$ of the $i$-th MP leads to a load force $f_l^i = \kappa_my^i$, where $\kappa_m$ is the force constant. Thus, the average load force on the motors is $f_l=\kappa_m y$ where $y = \frac{1}{n_m}\sum_{i=1}^{n_m}y^i$ denotes the average extension of these MPs. With $N_m$ MPs available on average and $n_m$ of them attached to the actin bundle at time $t$, the kinetics of the attached MPs is given by
\begin{gather}
	\frac{dn_m}{dt}={\omega_a}(N_m-n_m)-\omega_d^0n_m\exp{\left(\frac{|f_l|}{{f}_d}\right)}
	\label{eq:nmdyn}
\end{gather}

In their attached state, MPs move along the filament bundle with a velocity $v_m(f_l)$ which is dependent on the load force it experiences, predominantly towards one end of the filament. We model this behaviour with the piecewise linear force-velocity relation.
\begin{equation}
    v_m(f_l) = \begin{cases}
v_u  &\text{for $f_l \leq 0$}\\
v_u\left( 1 - \frac{f_l}{f_s}\right) &\text{for $0 < f_l \leq f_s$}\\
v_{b} &\text{for $f_l > f_s$}
\end{cases}
\label{eq:piecewise}
\end{equation}

where $f_s$ is the stall force when the MP ceases to move, $v_u$ is the intrinsic motor velocity without load and $v_b$ is a back velocity.

The motion of the MPs on the actin bundle induces the retrograde motion of the actin. The clutches are also modelled as extensible elastic springs with spring constant $\kappa_c$. One end of a clutch is attached to the actin bundle, while the other end is attached to an elastic substrate with stiffness $k_s$ (Fig.~\ref{fig:diagram}(b)). Retrograde motion of the actin bundle due to myosin contractility leads to an extension $x_c^i$ in the $i$-th attached clutch. The substrate extension, $x_s$, is determined by an elastic force balance between the total force due to the attached clutches and the spring force due to the substrate. We consider $N_c$ to be the total number of available clutches, $n_c$ as the number of clutches attached at a given time, and $x_c = \frac{1}{n_c}\sum_{i=1}^{n_c}x_c^i$ as the average extension of connected clutches. The dynamics of the average clutch deformation $x_c$ is determined by a mechanical balance of forces in the over-damped limit,
\begin{gather}
   \Gamma\frac{d{x}_c}{dt} = -n_m \kappa_my - n_c\kappa_cx_c
   \label{eq:xcdyn}
\end{gather}

where the viscous force due to motion of clutches, designated by viscous friction coefficient $\Gamma$, is balanced by the total restoring forces of both motors and clutches. Note that the negative sign in the force expression for the clutches in the above equation is a matter of convention since $x_c$ takes negative values. Within our assumption of treating the motor-clutch sector and substrate sector as separate blocks, the deformation of the substrate can be calculated independently using the force balance $k_s x_s = - n_m \kappa_my - n_c \kappa_c x_c$. This allows us to focus on the stability of the motor-clutch sector alone.

The rate of mean extension of an attached MP is determined by the active motor velocity $v_m$ on the filament and the rate of average deformation of the attached clutches. This is given as
\begin{gather}
	\frac{d{y}}{dt} = v_m(f_l)+\frac{d{x}_c}{dt}
	\label{eq:ydyn}
\end{gather}

Clutches undergo attachment-detachment dynamics with rates $k_{\text{on}}$ and $k_{\text{off}}$ respectively. The clutch detachment rate is again assumed to be exponentially increasing with the load force, $k_{\text{off}} = k_{\text{off}}^0\exp{(|f_l^c|/F_b)}$, where $F_b$ is the force scale for bond rupture. Similar to the MPs, the average load force on clutches is $f_l^c=\kappa_c x_c$. The attachment-detachment dynamics of the clutches gives rise to the following rate equation. 
\begin{gather}
  \frac{dn_c}{dt}=k_{\text{on}}(N_c-n_c)-k_{\text{off}}^0n_c\exp\left(\frac{|f_l^c|}{F_b}\right)
  \label{eq:ncdyn}
\end{gather}

\begin{table}
    \centering
      \begin{tabular}{lcr}
      \toprule
      Parameter & Symbol & Values\\
      \midrule
      Motor attachment rate & $\omega_a$ & $40\,\text{s}^{-1}$~\cite{walcott2012mechanical}\\
      Motor detachment rate & $\omega_d^0$ & $350\,\text{s}^{-1}$~\cite{walcott2012mechanical}\\
      Total number of motors & $N_m$ & 100 \\
      Clutch attachment rate & $k_\text{on}$ & $1\,\text{s}^{-1}$~\cite{chan2008traction}\\
      Clutch detachment rate & $k_\text{off}$ & $0.1\,\text{s}^{-1}$~\cite{chan2008traction}\\
      Back velocity & $v_b$ & $0.2256\,\mu\text{m s}^{-1}$~\cite{walcott2012mechanical}\\
      Stall force & $f_s$ & 4.96 pN~\cite{walcott2012mechanical}\\
      Detachment force & $f_d$ & 2.4 pN~\cite{schnitzer2000force}\\
      Clutch bond rupture force & $F_b$ & 6.25 pN~\cite{bangasser2013determinants}\\
      Motor spring constant & $\kappa_m$ & 0.3 pN/nm~\cite{walcott2012mechanical}\\
      Clutch spring constant & $\kappa_c$ & 0.03144 pN/nm~\cite{dembo1988reaction}\\
      Viscous friction coefficient & $\Gamma$  & $893 k_B$Ts/{$\mu$}m$^2$~\cite{Lansky2015}\\
      \bottomrule
      \end{tabular}
      \caption{Physical parameters present in the system}
      \label{tab:parameters}
\end{table}
We present a detailed analysis of the stability of the MP-filament-clutch system emphasising the effect of the activity of myosin motors both in terms of the motor velocity and attachment/detachment kinetics. The physical parameters used in our model are described in the Table \ref{tab:parameters}. Choosing the length, time, velocity and force scales as $l_0=({k_bT}/{\omega_d^0\Gamma})^{1/2}$, $\tau = 1/\omega_d^0$, $v_0=l_0\omega_d^0$, and $f=(\omega_d^0\Gamma k_bT)^{1/2}$, respectively, Eqs.\eqref{eq:nmdyn}, \eqref{eq:xcdyn}, \eqref{eq:ydyn}, and \eqref{eq:ncdyn} are cast in dimensionless form with $\tilde\omega={\omega_a}/{\omega_d^0}$, $\tilde{v}_u = v_u/v_0$, $\tilde{k}_{\text{on}}=k_{\text{on}}/{\omega_d^0}$, $\tilde{k}_{\text{off}}=k_{\text{off}}/{\omega_d^0}$, $\tilde{x}_c={x_c}/{l_0}$, $\tilde{y}={y}/{l_0}$, $\tilde{\kappa}_c={\kappa_cl_0}/{f}$, $\tilde{\kappa}_m={\kappa_ml_0}/{f}$, and $\tilde{f}_s={f_s}/{f}$ (see \hyperref[s:appxa]{Appendix A}). Attachment/detachment dynamics is varied using a turnover ratio defined as $\Omega = \omega_a/(\omega_a + \omega_d^0)$. A dynamical modelling of the system provides us with the basic building blocks of understanding the mechanics of motility in the absence of noise. We proffer a \textit{linear stability analysis} of the system, numerical solutions of the differential equations, illustrate the morphologies and characterise the detailed dynamics.

\subsection*{Stochastic simulation of motor-clutch system}
We model the actin filament as a rigid string of connected $\sigma = 5.5$ nm segments~\cite{stam2015isoforms,veigel2003load}. The \(i\)th myosin motor can attach to an actin segment stochastically with the rate $\omega_a$. At the moment of attachment, the extension of the motor protein, $y^{i}$, is zero. Post attachment, the MP moves by a length scale $\sigma$ towards the plus (minus) end of the actin filament with a velocity $v_m$ where $v_m$ is given by Eq.~\eqref{eq:piecewise}. The total extension of the attached MP, $y^i$, is determined by the active motor velocity $v_m$ on the filament and the deformation of the attached clutches. The MPs detach from the actin filament with a rate $\omega_d = \omega_d^0\exp{(\kappa_m|y^i|/f_d)}$.

The $i$th clutch undergoes attachment and detachment dynamics with rates $k_{\textrm{on}}$ and $k_{\textrm{off}} = k_{\textrm{off}}^0\exp{(\kappa_c|x_c^i|)}$, respectively. After attachment, the clutch deformation \(x_c^i\) is determined from the following stochastic equation
\begin{equation}
\Gamma \dot{x}_c^i = -\sum_{i=1}^{n_m}\kappa_my^i - \sum_{i=1}^{n_c}\kappa_cx_c^i + \sqrt{2\Gamma k_BT}\eta_T(t)
\end{equation}
as opposed to Eq.~\eqref{eq:xcdyn}. Here $\eta_T(t)$ is a Gaussian noise with $\langle \eta_T(t) = 0\rangle$ and $\langle \eta_T(t) \eta_T(t^{\prime})\rangle = \delta(t - t^{\prime})$. We numerically integrate the above equation using Euler-Maruyama scheme with time steps small enough that the probability of each event is less than one. The stochastic simulations were performed using programs written in Fortran.

\section*{Results}
The steady state solutions, i.e the fixed points $\tilde{x}^0_c$, $\tilde{y}_0$, $n_m^0$, $n_c^0$, of the scaled dynamical equations are obtained as $\tilde{y}_0 = \tilde{f}_s/\tilde{\kappa}_m$, $n_m^0 = \tilde\omega N_m/({\tilde\omega+\exp{(\tilde{f}_s}/\tilde{f}_d)})$, and $\tilde{x}^0_c = -n_m^0\tilde{f}_s/n_c^0\tilde{\kappa}_c$. Thus, the extension of the clutches in steady state is governed by the ratio of the numbers of attached motors and attached clutches. $n_c^0$ is determined by solving the transcendental equation,
\begin{align}
\tilde{k}_{\text{on}}(N_c-n_c^0)&=\tilde{k}_{\text{off}}n_c^0\exp{\left(\frac{n_m^0 \tilde{f}_s}{n_c^0\tilde{F}_b}\right)}
\label{eq:trans_nc0}
\end{align}

This equation, as we shall see, gives rise to a saddle-node bifurcation with two branches -- one stable and another unstable.

The stability of these fixed points is tested by studying the time evolution of small perturbations away from the steady state. In that vein, the dynamical equations representing the system in terms of MP extension, clutch extension, number of attached motors, and number of connected clutches can be linearised in matrix form, $\frac{d}{d\tau}[\delta\tilde{x}_c,\delta\tilde{y},\delta n_m,\delta n_c]^T=\underbar{\rm {\bf J}}[\delta\tilde{x}_c,\delta\tilde{y},\delta n_m, \delta n_c]^T$. Eigenvalues of the $4\times 4$ Jacobian matrix $\underbar{\rm {\bf J}}$, (see \hyperref[s:appxb]{Appendix B}) determine the linear stability of the dynamical system. The eigenvalues are calculated by solving the fourth-order characteristic polynomial equation,
\begin{align}
	P(\lambda) = \lambda^4+\mathcal{A}\lambda^3+\mathcal{B}\lambda^2+\mathcal{C}\lambda+\mathcal{D}
	\label{eq:eigen_quartic}
\end{align}

where $\mathcal{A, B, C}$ and $\mathcal{D}$ are the coefficients which are given in terms of the scaled parameters (see \hyperref[s:appxb]{Appendix B}).

The nature and properties of the eigenvalues are dependent on the sign of the coefficients $\mathcal{A}$, $\mathcal{B}$, $\mathcal{C}$ and $\mathcal{D}$. We note that $\mathcal{A}$ is always real positive, however the other three coefficients can change signs and different combinations of those signed coefficients determine the nature of the roots of the quartic polynomial and related dynamical phases. With $\lambda_k$ ($k = 1\ldots4$) denoting the eigenvalues, we get the following combination: (a) all real negative eigenvalues corresponding to stable nodes where a perturbation decays exponentially with time; (b) 2 real negative and 2 real positive - both these combinations correspond to the unstable phase with exponentially growing perturbations; (c) 2 real negative and 2 eigenvalues with $\lambda_{3,4} = -\alpha\pm i\beta$ resulting in stable spiral phase with decaying oscillations; and (d) 2 real negative and 2 eigenvalues with $\lambda_{3,4} = \alpha\pm i\beta$ corresponding to the unstable spiral phase with growing oscillations. We proceed with determining the closed-form expressions for the possible phase boundaries present in our system. \textit{Wolfram Mathematica}~\cite{Mathematica} was used to numerically solve the dynamical equations using \textit{Implicit Differential-Algebraic Solver}~\cite{IDA}.
\begin{figure}
    \centering
    \includegraphics{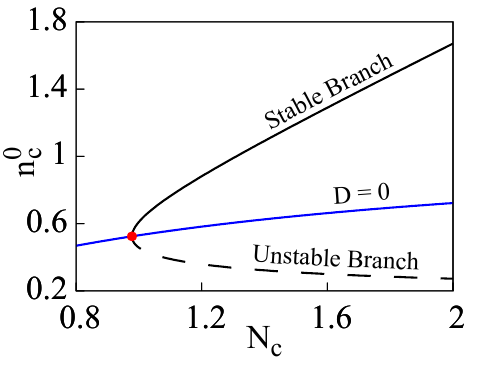}
    \caption{Transcendental nature of the $n_c^0$ equation leads to a \textit{saddle-node bifurcation} with distinguishable branches -- solid black curve representing the stable branch and dashed black line denoting the unstable one. The red point denotes the bifurcation point of the system.}
    \label{fig:bifurcation}
\end{figure}

\subsection*{Phase boundary separating saddle-node bifurcated stable and unstable branches}
The stability of the two branches arising from a saddle-node bifurcation in Eq.~\eqref{eq:trans_nc0} can be characterised by checking the sign of its derivative. A negative/positive value will represent a stable/unstable branch of fixed points and the \textit{bifurcation point} can be obtained by solving Eq.~\eqref{eq:trans_nc0} while simultaneously setting its derivative with respect to $n_c^0$ to zero. The derivative equation is computed below,
\begin{align}
  -\tilde{k}_\text{on} - \left[\tilde{k}_\text{off}\exp{\left(\frac{n_m^0 \tilde{f}_s}{n_c^0 \tilde{F}_b}\right)} - \tilde{k}_\text{off} \frac{n_m^0 \tilde{f}_s}{n_c^0 \tilde{F}_b}\exp{\left(\frac{n_m^0 \tilde{f}_s}{n_c^0 \tilde{F}_b}\right)}\right] = 0
\end{align}

\begin{figure*}[t!]
  \centering
  \includegraphics[width=\textwidth]{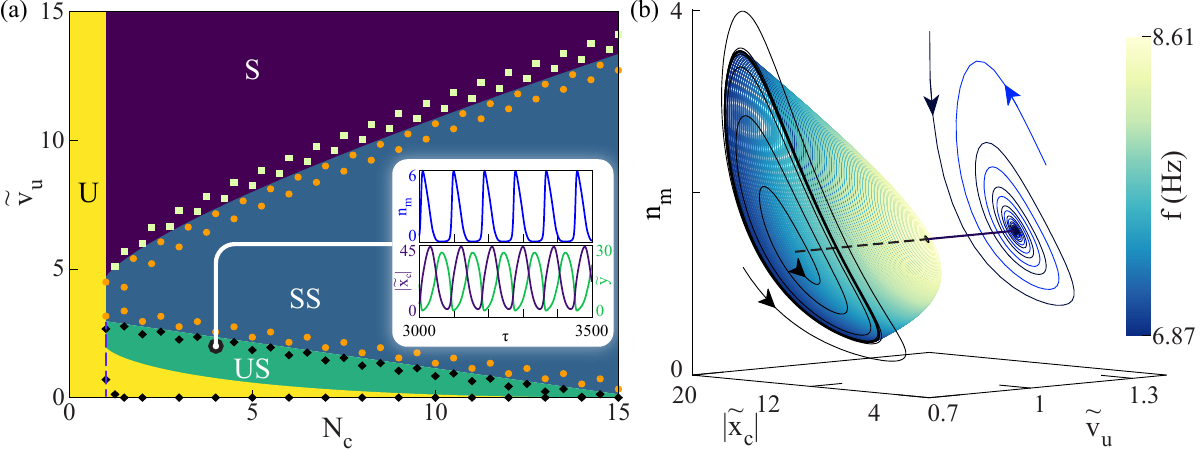}
  \caption{(a) \textbf{Phase diagram depicting dynamical phases in the $N_c$--$\tilde{v}_u$ plane:} Phase boundaries predicted by stability analysis and the resultant dynamical phases are portrayed using coloured regions with labels --  \colorbox[HTML]{440154}{\color{white}{S}} stable, \colorbox[HTML]{33638D}{\color{white}{SS}} stable spirals, \colorbox[HTML]{29AF7F}{\color{white}{US}} unstable spirals, and \colorbox[HTML]{FDE725}{{U}} unstable. The unstable region consists of two parts -- one arising from an absence of fixed points and another from the loss of stability due to $C<0$. A dashed line is used to separate them. Coloured points are obtained by solving the dynamical equations numerically near the phase boundaries  -- {\color[HTML]{DFFFAD}{$\blacksquare$}} stable, {\color[HTML]{FF9D00}{\textbullet}} decaying oscillations, and $\blacklozenge$ limit cycle oscillations. The range of $\tilde{v}_u$ is equivalent to $0\sim 9~\mu$m/s, in physical units. The phase boundary demarking Hopf bifurcation, i.e. between \colorbox[HTML]{29AF7F}{\color{white}{US}} and \colorbox[HTML]{33638D}{\color{white}{SS}}, is provided by Eq.~\eqref{eq:pbhopf}, while the boundaries between U/US and SS/S are governed by Eq.~\eqref{eq:pbnosp}. (inset) \textbf{Time evolution.} At $N_c=4$ and $\tilde{v}_u=2$, the system shows stable limit cycle oscillations and corresponding evolution of $n_m$, $\tilde{y}$ and $|\tilde{x}_c|$ are shown in panels. The ranges of $|\tilde{x}_c|$ and $\tilde{y}$ are equivalent to $0\sim80$ nm and $0\sim53$ nm respectively, in physical units. The time $t$ ranges from 8.5 to 10 seconds. (b) \textbf{Super-critical Hopf bifurcation:} At $N_c = 10$, the system shows decaying oscillations at higher values of $v_u$ and stable limit cycle oscillations through a super-critical Hopf bifurcation as $\tilde{v}_u$ is lowered. The straight line shows the fixed points in the $n_m-|\tilde{x}_c|$ plane. Solid(dashed) line denotes their stable(unstable) nature. As the limit cycle grows with decreasing $\tilde{v}_u$, its frequency starts to reduce. Here $v_u$ and $|x_c|$ range from $0.4\sim0.8 \mu$m/s and $7\sim35$ nm respectively, in physical units.}
  \label{fig:phased}
\end{figure*}

With trivial algebra, it can be easily shown that this is exactly the same as the phase boundary equation at $D = 0$,
\begin{align}
  n_c^0\tilde{k}_{\text{on}}\tilde{F}_b+\tilde{k}_{\text{off}}\exp{\left(\frac{n_m^0 \tilde{f}_s}{n_c^0 \tilde{F}_b}\right)}\left(\tilde{F}_b n_c^0-\tilde{f}_s n_m^0\right) = 0
  \label{eq:perpunspb}
\end{align}
The bifurcation point is then calculated and marked in the Fig.~\ref{fig:bifurcation} with a red dot. Using Eq.~\eqref{eq:trans_nc0} one can simplify Eq.~\eqref{eq:perpunspb} and attain the following,
\begin{align}
  N_cn_c^0\tilde{F}_b = n_m^0\tilde{f}_s(N_c-n_c^0)
\end{align}
Confirming the presence of an unstable branch in the $\mathcal{D}<0$ region, we shift our focus to the stable branch where $\mathcal{D}>0$ and investigate the geometric properties of the quartic polynomial with the change in sign of $\mathcal{B}$ and $\mathcal{C}$. In the region where $N_c$ is lower than its value at the bifurcation point, due to the absence of fixed points of $n_c^0$ the system loses stability and remains unstable regardless of $v_u$. Therefore, this region is marked unstable on the phase diagram in Fig.~\ref{fig:phased}(a). Furthermore, the condition $B>0$ remains valid throughout the scanned parameter regime. Consequently, the properties of coefficient $\mathcal{C}$ associated with the linear part of the polynomial $P(\lambda)$ mostly governs the nature of the eigenvalues.

\subsection*{Phase boundary between (un)stable nodes and (un)stable spiral phases}
The quartic polynomial $P(\lambda)$ is bound from below as the term $\lambda^4$ comes with positive sign and has got two minima and one maximum. Sign of $\mathcal{C}$ controls the position of the minimum closer to the origin as $\mathcal{B}$ remains positive. When $P_m$, a minimum of $P(\lambda)$ occurring near the origin, with $\lambda=\lambda_m$, crosses the negative $\lambda$ axis (positive $\lambda$ axis), it turns two complex conjugate eigenvalues with negative (positive) real parts into real negative (real positive). Therefore, $P_m = 0$ denotes a boundary between phases with either a node or a spiral (refer to Fig.~\ref{fig:phased}), where $C>0$ ($C<0$) provides a sufficient condition for concluding that these fixed points are stable(unstable). A necessary and sufficient condition can be obtained by finding the equation of a second phase boundary, which can be derived from the fact that we have two degenerate real roots $\lambda_m$ at the boundary. Thus comparing the coefficients of various powers of $\lambda$ with those in Eqn. \ref{eq:eigen_quartic}, we obtain a closed-form equation of the stability boundary,
\begin{align}
2\mathcal{AD}\left(-9 \mathcal{A}^2 \mathcal{B} \mathcal{C}+2 \mathcal{A} \mathcal{B}^3+3 \mathcal{A} \mathcal{C}^2+40 \mathcal{B}^2 \mathcal{C}\right)+\mathcal{D}^2 \left(27 \mathcal{A}^4-144 \mathcal{A}^2 \mathcal{B}+192 \mathcal{AC}+128 \mathcal{B}^2\right)\nonumber\\
+\mathcal{C}^2 \left(4 \mathcal{A}^3 \mathcal{C}-\mathcal{A}^2 \mathcal{B}^2-18 \mathcal{A B C}+4 \mathcal{B}^3+27 \mathcal{C}^2\right)=16 \mathcal{D} \left(\mathcal{B}^4+9\mathcal{B C}^2+16 \mathcal{D}^2\right)
\label{eq:pbnosp}
\end{align}
\begin{figure*}[t!]
    \includegraphics{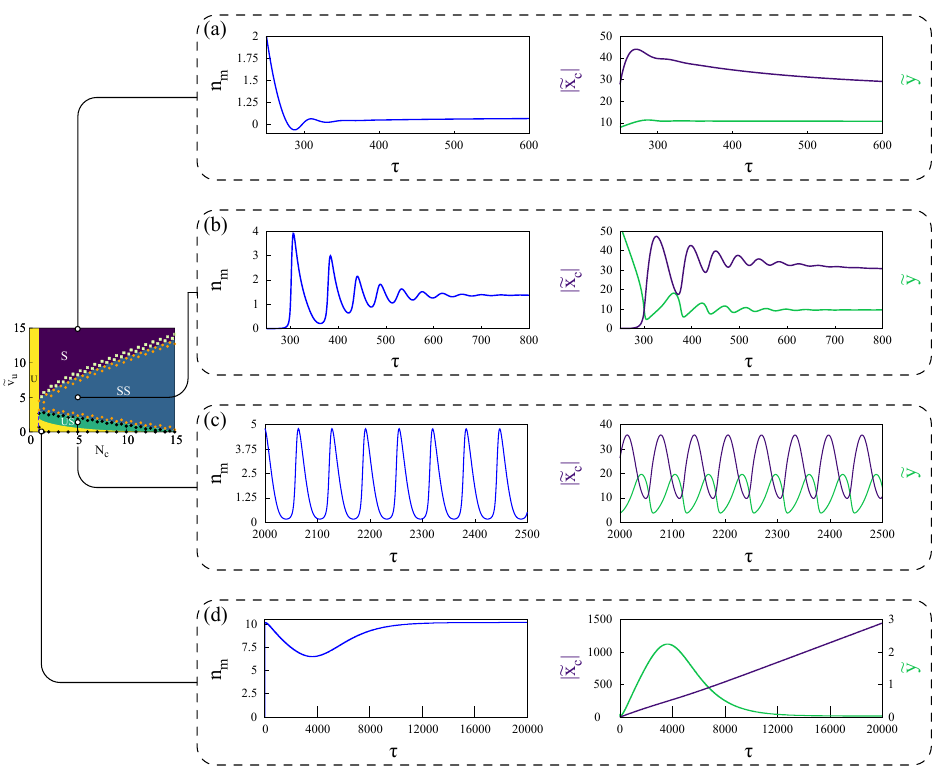}
    \caption{\textbf{Dynamics of the active system}: Numerical solutions of the dynamical equations at different points in the $N_c-\tilde{v}_u$ phase plane elucidate the dynamical phases predicted by \textit{linear stability analysis}. The range of $\tilde{v}_u$ is equivalent to $0\sim 9~\mu$m/s, in physical units. At high $\tilde{v}_u$ region (a), the system is quickly stabilised as it reaches steady-state solutions, while relatively lower $\tilde{v}_u$ at (b) ensures that the system follows a path of decaying oscillations. It is at (c), where the system showcases self sustaining limit cycle oscillations after crossing the supercritical Hopf bifurcation boundary. As predicted by linear stability analysis, critically low values of $\tilde{v}_u$ and $N_c$ ensure that the system is unstable, as evident by runaway clutch deformation. The range of extensions are equivalent to $0$ -- $70\sim88$ nm in the first three plots. In (d), $x_c$ ranges within $0\sim2650$ nm, while $y$ remains between $0\sim5$ nm. Ranges of time in physical units are $0.6\sim1.7$ seconds in (a), $0.6\sim2.3$ seconds in (b), $5.7\sim7.1$ seconds in (c), and $0\sim57$ seconds in (d).}
    \label{fig:pbco}
\end{figure*}
\subsection*{Phase boundary between stable spirals and unstable spirals}
An oscillation increasing in time appears in the system, switching from decaying oscillation as the parameter values are tuned. It is characterised by a change in complex conjugate eigenvalues from $(-\alpha\pm i\beta)$ to $(\alpha\pm i\beta)$ that is an occurrence of a dynamical transition between stable spiral(SS) and unstable spiral(US). The sign of the real part of the complex conjugate roots $\alpha$ is opposite in either sides of the boundary and thereupon, $\alpha = 0$ is the condition of the associated phase boundary. Following the prescription of the previous case, we calculate the equation of the boundary as,
\begin{align}
  \mathcal{ABC} = \mathcal{A}^2\mathcal{D}+\mathcal{C}^2
  \label{eq:pbhopf}
\end{align}
SS to US transition is the route through which stable limit cycle oscillation sets in the system via non linear effects leading to {\em Hopf bifurcation}. At the bifurcation boundary, we can work out an expression for the frequency as $f_{\omega} = (2\pi)^{-1}\sqrt{\mathcal{C}/\mathcal{A}}$.
\begin{figure}
	\centering
	\includegraphics{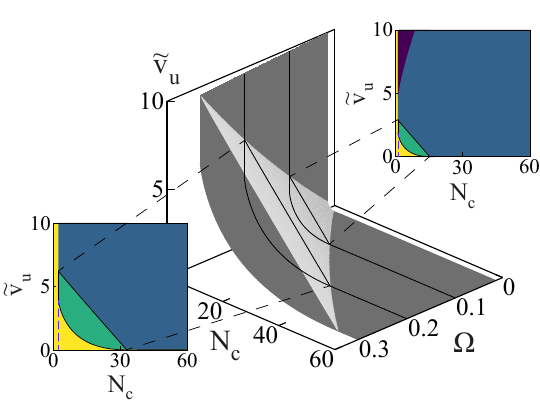}
	\caption{\textbf{Evolution of phase boundaries with varying active velocity $\tilde{v}_u$ and myosin turnover $\Omega$:} Increasing both the active velocity and motor attachment rates result in growth of areas enclosed by the phase boundaries,  namely the boundaries separating U and US phases (dark gray) and the Hopf bifurcation boundaries (light gray). We focus on two slices obtained at duty ratios 0.1 and 0.2 respectively. The enclosed areas (green) in the resulting phase diagrams clearly establish this expansion. The range of $\tilde{v}_u$ is equivalent to $0\sim 6~\mu$m/s, in physical units.}
	\label{fig:turnover}
\end{figure}
\subsection*{Force fluctuations: regulatory pathway via Hopf bifurcation}
To quantify the explicit effect of myosin activity, we present a phase diagram in $\tilde{v}_u$ and $N_c$ plane which illustrates the different dynamical behaviours of the motor-clutch system (Fig.~\ref{fig:phased}(a)). As observed earlier, $N_c$ controls the saddle-node bifurcation and $v_u$ regulates the activity of the myosin motors, providing an apt parameter space for the model mechanics. Earlier experimental studies ~\cite{HarrisWarshaw,Debold,Debold2011} have established myosin force-velocity relations and ensemble measurements with unloaded motor velocities. We used the broad range of $0 - 10 \mu$m/s for $v_u$ in our study. These predictions are evaluated by the numerical solutions of the differential equations over the entire parameter space. The phase boundaries predicted by the \textit{linear stability analysis} are exact. At high motor velocities, the system is in a stable state. As the motor velocity is reduced, the system moves into a stable spiral state, as seen by the inward spiralling curve in the $n_m - x_c$ phase plane (Fig.~\ref{fig:phased}(b)). With sufficient number of clutches available, as $\tilde{v}_u$ is lowered, the oscillations in stable spiral region take a gradually increasing amount of time to decay and cross into a stabilised, oscillatory region via a super-critical Hopf bifurcation. This indicates a limit cycle around an unstable fixed point in the $n_m - x_c$ plane. The temporal oscillations of the clutch and motor deformations in the unstable spiral regime are shown in Fig.~\ref{fig:phased}(a)(inset). At sufficiently low motor velocities, $\tilde{v}_u$, the system moves from a region of instability to unstable spiral on increasing $N_c$.

The physical understanding of the stability mechanism in the motor clutch system is achieved via two balancing acts: stalling of motors with a particular extension given by $\tilde{k}_m\tilde{y}_0=\tilde{f}_s$ and the force balance  $n_{m}^0\tilde{f}_s=n_{c}^0\tilde{k}_c(-\tilde{x}_c^0)$. The modalities of this mechanism are corroborated by the numerical solutions of the dynamical equations across a range of active velocities in Fig.~\ref{fig:pbco}. At high ATP concentrations, motor velocities ($v_u$) are high (Fig.~\ref{fig:pbco}(a)), leading to large extensions and hence stalling of motors is quickly established. The clutch extension attains the overall force balance condition, and the system becomes stable. With higher number of clutches, stability is attained at higher $\tilde{v}_u$. At low ATP concentrations (Fig.~\ref{fig:pbco}(b-c)), with slower moving motors, the forces exerted by motors on the actin filaments results in slow retrograde movement of the filament which in turn reduces motor deformation $\tilde{y}$ and increases clutch extension $\tilde{x}_c$. Reduction in $\tilde{y}$, results in an increase in the number of motors ($n_m$) with time resulting in further retrograde movement of the actin filament. The leftward movement of the filament is stopped when $n_m$ reaches its maximum, whereas $\tilde{x}_c$ and $n_c$ are at their near maximum and minimum respectively. Beyond this point, slow motor velocity results in detachment of motors, and the stored energy in the deformed clutch sector ensures the anterograde movement of the filament. Motor detachment continues, and the filament moves towards less and less clutch deformation until it reaches a minimum. $\tilde{y}$ attains maximum and the motor attachment rate takes over its detachment rate as $n_m$ becomes minimum. The cycle continues. At very low ATP concentrations, with the motor velocities ($v_u$) very small (Fig.~\ref{fig:pbco}(d)), the deformation $\tilde{y}$ continues to go down with the retrograde movement of the actin filament. However, after reaching force balance, the extremely slow movement of the motors means that the motor extension beyond the point is never realized. The clutches on the other hand are at their highest extension and with no relief, fail completely which leads to instability. 

In Fig.~\ref{fig:turnover}, we vary myosin turnover, $\Omega$, by tuning the attachment rates of myosin. This provides another experimentally tunable mechanism of changing myosin contractility. The phase diagram shows that with increasing $\Omega$, the phase boundaries between the unstable (U) - unstable spiral (US) and unstable spiral (US) - stable spiral (SS) shifts towards larger $\tilde{v}_u$ and $N_c$, and the area enclosed between them expand. This signifies that the motor proteins which have a higher tendency to attach will also result in more persistent limit cycles over larger areas in the parameter space.

\begin{figure}
  \includegraphics[width=\textwidth]{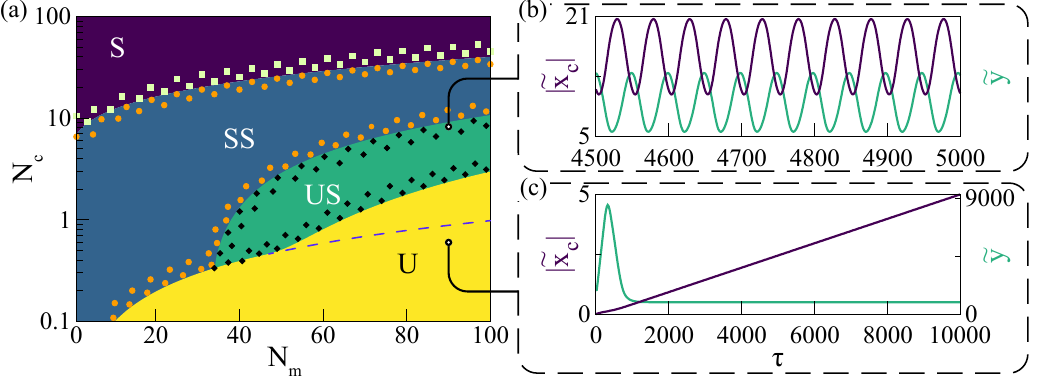}
  \caption{\textbf{Phase diagram in the $\mathbf{N_m-N_c}$ plane:} Tuning total numbers of motors and clutches simultaneously at a fixed $\tilde{v}_u = 1$ leads to a phase diagram with dynamical phases discussed before -- \colorbox[HTML]{440154}{\color{white}{S}} stable, \colorbox[HTML]{33638D}{\color{white}{SS}} stable spirals, \colorbox[HTML]{29AF7F}{\color{white}{US}} unstable spirals, and \colorbox[HTML]{FDE725}{{U}} unstable. A dashed line is used to separate the two unstable regions discussed previously. The coloured points are used following the convention from Fig.~\ref{fig:phased} to test the robustness of the phase boundaries predicted by linear stability analysis. \textbf{Time evolution} of the system at two different points are depicted in (b). The upper point is below the Hopf bifurcation boundary and exhibits limit cycle oscillations as expected, while the lower point is well inside the unstable region and a runaway $|\tilde{x}_c|$ establishes the predicted instability. The length scales in (b) and (c) range between 8.8 -- 37 nm and 0 -- 8.8 nm, respectively. Time scales for these figures range between 12.9 -- 14.2 seconds and 0 -- 28.6 seconds, respectively.}
  \label{fig:phasednmnc}
\end{figure}
Examining our system with tunable total motor/clutch numbers allows us to probe the dynamic behaviour under another important experimentally viable parameter space. Linear stability analysis in this space leads to the same array of dynamic phases seen before. We also test the robustness of predicted phase boundaries by incorporating numerical solutions to the differential equations, shown in Fig.~\ref{fig:phasednmnc}~(a). Time evolutions of motor and clutch extensions at two points of interest on the phase diagram are portrayed in the accompanying plots in Fig.~\ref{fig:phasednmnc}~(b). We indeed find that for a given value of the motor velocity, there is a minimum number of motors and clutches required to observe the oscillations.

\begin{figure*}
    \centering
    \includegraphics[width=\textwidth]{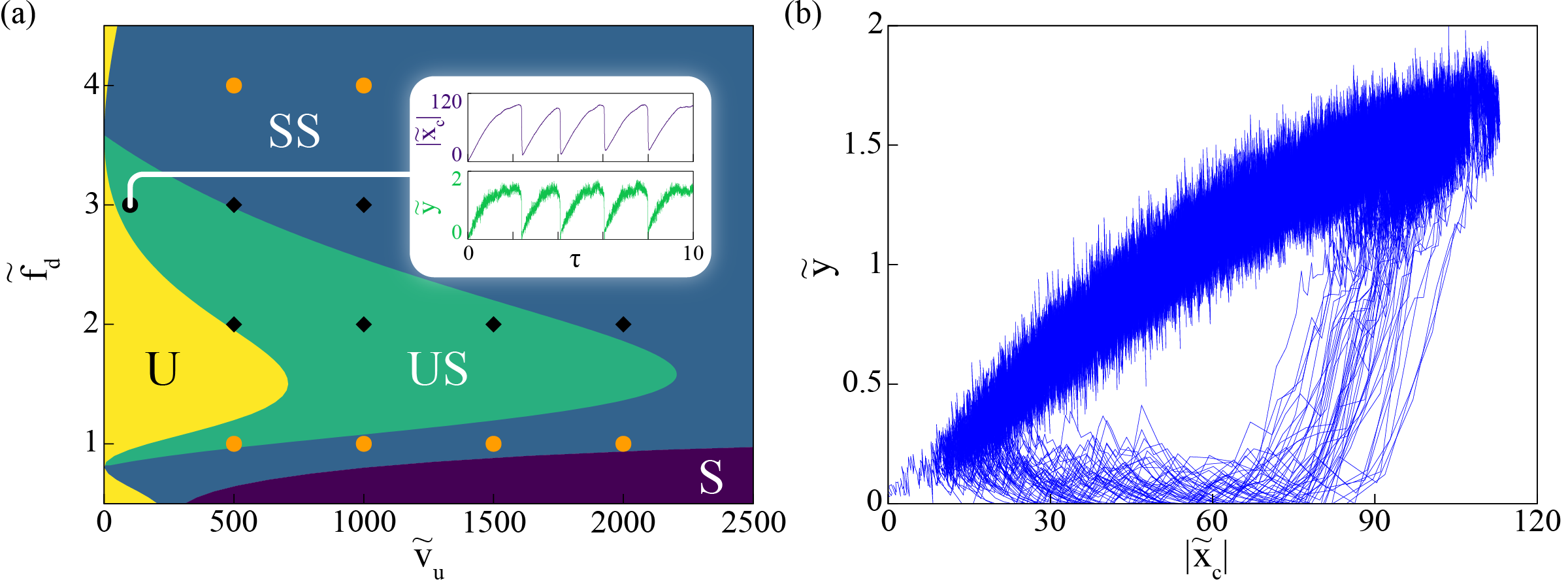}
    \caption{(a) \textbf{Phase diagram portraying permanently bound state of clutches} ($k_{\textrm{off}} = 0$) in the $\tilde{v}_u - \tilde{f_d}$ plane, for fixed $\tilde{\omega} = 5, \tilde{f}_s = 6.58815, N_m = 100, N_c = 10, \tilde{\kappa}_m = 2.191$ and $\tilde{\kappa}_c = 0.2296$. Following the previous convention, coloured regions with labels indicate -- \colorbox[HTML]{440154}{\color{white}{S}} stable, \colorbox[HTML]{33638D}{\color{white}{SS}} stable spirals, \colorbox[HTML]{29AF7F}{\color{white}{US}} unstable spirals, and \colorbox[HTML]{FDE725}{{U}} unstable dynamical phases. Stochastic simulations were run at the points, confirming two phases characterising {\color[HTML]{FF9D00}{\textbullet}} decaying oscillations, and $\blacklozenge$ limit cycle oscillations. In physical units, $v_u$ and $f_d$ range between 0 -- 2.75 $\mu$m/s and 0 -- 3.75 pN, respectively. (inset) \textbf{Time evolution} At $\tilde{v}_u = 100, \tilde{f_d} = 3$, the system produces a stable limit cycle and the analogous stochastic time evolution of clutch displacement $|\tilde{x}_c|$ and mean MP extension $\tilde{y}$ are presented in the panel. In physical units, $|x_c|$ and $y$ range between 0 -- 660 nm and 0 -- 11 nm, respectively. (b) \textbf{Limit cycle} A parametric plot of the $|\tilde{x}_c(\tau)|$ and $\tilde{y}(\tau)$ showing stable limit cycle oscillations. The length scales here are same as in the inset figures.}
    \label{fig:trap}
\end{figure*}

\subsection*{Simulation output validates motor-clutch model}
In order to check our results, we first consider the special case where all clutches are attached permanently i.e $k_{\textrm{off}} = 0$. To present a comparison between the theoretical model and the numerical simulations,  we first present the results from the dynamical equations.  We now have three coupled differential equations in the number of attached motors ($n_m$), the average deformation of a clutch ($x_c$) and the average deformation of a molecular motor ($y$).  As in Section 3, we obtain the steady state solution of the coupled scaled differential equations as $\tilde{y}_0 = \tilde{f}_s/\tilde{\kappa}_m$, $n_m^0 = \tilde\omega N_m/({\tilde\omega+\exp{(\tilde{f}_s}/\tilde{f}_d)})$ and $\tilde{x}^0_c = -n_m^0\tilde{f}_s/N_c\tilde{\kappa}_c$.  The transcendental equation in $n_c^0$ is now replaced by the constant $N_c$, as all the clutches are now bound.  We perform a linear stability analysis by studying the time evolution of small perturbations away from the steady state. This leads to a third-order characteristic polynomial equation in the eigenvalues $\lambda$ as $P^{\prime}(\lambda) = \lambda^3 + {\cal A}^{\prime}\lambda^2 + {\cal B}^{\prime}\lambda + {\cal C}^{\prime}$, where ${\cal A}^{\prime}, {\cal B}^{\prime}$ and ${\cal C}^{\prime}$ are the new coefficients given in terms of scaled parameters (see \hyperref[s:appxd]{Appendix D} for details).

The coefficients determine the dynamical behaviour of the system which has four different phases characterised by the different combinations of the three eigenvalues: (1) All three eigenvalues real negative which result in stable nodes (2) 1 negative and 2 real positive giving rise to a linearly unstable phase (3) 1 real negative and two complex conjugates with negative real parts, characterising a stable spiral phase with decaying oscillations and (4) 1 real negative and two complex conjugate roots with positive real parts characterising an unstable spiral phase with oscillations of growing amplitude. We can determine the different phase boundaries analytically as before. However, to show the comparison with numerical simulations which incorporate stochasticity as described above, we concentrate on the phase boundary between the stable spiral and unstable spiral phases. The condition for the phase boundary is ${\cal C}^{\prime} - {\cal A}^{\prime}{\cal B}^{\prime} = 0$. Using numerical simulations we show how the growing amplitudes of the oscillations in the unstable spiral phase are stabilised by non-linearities into stable limit cycle oscillations.

In Fig.~\ref{fig:trap}(a), we present the phase diagram of the system in the $\tilde{v_0}-\tilde{f_d}$ plane. As we can observe, there are four phases, with two being of particular interest: (1) corresponding to the stable spiral phase \colorbox[HTML]{33638D}{\color{white}{SS}} characterised by decaying oscillations, and (2) corresponding to the unstable spiral phase \colorbox[HTML]{29AF7F}{\color{white}{US}} characterised by stable limit cycle oscillations. We also plot the phase boundary as obtained from our analytical estimates which shows reasonable agreement with the simulations. In inset of Fig.~\ref{fig:trap}(a), we show the dynamical behaviour of the scaled clutch deformation $\tilde{x_c}$ and the scaled motor extension $\tilde{y}$ in the stable limit cycle phase. The slow extension and the rapid decay of the clutch deformation is reminiscent of the rapid detachment of MPs. The parametric plot in Fig.~\ref{fig:trap}(b), shows a stable limit cycle as expected. The spread in the trajectories underlines the stochastic nature of the simulations.

\begin{figure*}
  \centering
  \includegraphics[width=\textwidth]{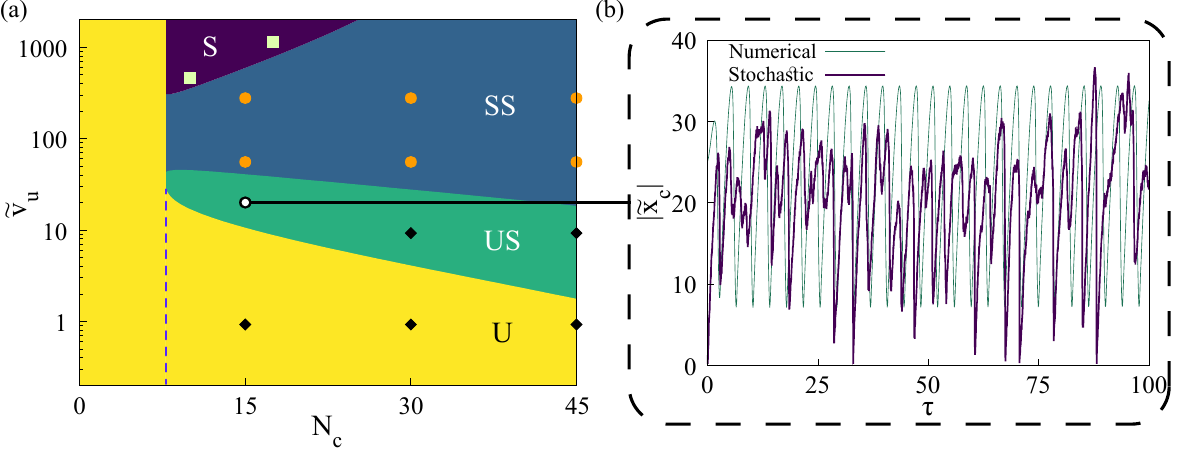}
\caption{(a) \textbf{Phase diagram depicting dynamical phases in the $N_c - {\tilde{v}_u}$ plane} for fixed $\tilde{\kappa}_m = 2.191, \tilde{\omega} = 1, \tilde{f}_s = 6.58815, N_m = 100$ and $\tilde{\kappa}_c = 0.2296$. Convention-wise, coloured regions with labels indicate -- \colorbox[HTML]{440154}{\color{white}{S}} stable, \colorbox[HTML]{33638D}{\color{white}{SS}} stable spirals, \colorbox[HTML]{29AF7F}{\color{white}{US}} unstable spirals, and \colorbox[HTML]{FDE725}{{U}} unstable dynamical phases. A dashed line is used to separate the two unstable regions discussed previously. Stochastic simulations at the indicated points confirmed the existence of {\color[HTML]{DFFFAD}{$\blacksquare$}} stable phase, {\color[HTML]{FF9D00}{\textbullet}} decaying oscillations, and $\blacklozenge$ limit cycle oscillations. (b) \textbf{Comparison of limit cycle oscillations} in the clutch deformation from numerical solution of the differential equations and the stochastic simulations. In physical units, $x_c$ ranges from 0 -- 220 nm.}
\label{fig:withkoff}
\end{figure*}

Having established the simulation model, we advanced to verify the results for the situation where the clutches are free to attach/detach to/from the actin filament. In Fig.~\ref{fig:withkoff}(a), we show the phase diagram obtained using the equations as described in Section 2. As before, we see the five different phases with different phase boundaries. In Fig.~\ref{fig:withkoff} (b), we show a comparison of the limit cycle oscillations between the theoretical calculations and numerical simulations incorporating stochasticity arising from different sources - attachment/detachment of MPs and clutches to/from the actin filament, stochastic extension of attached clutches and MPs and finally the Gaussian noise. As we can observe, we do recover stable limit cycle oscillations in the given parameter regime as predicted by our theoretical study.

\section*{Discussion}
The variability of cell traction force measurements suggests that a mere readout of these forces may not be optimal in understanding the processes that regulate force generation and subsequent transmission~\cite{kurzawa2017dissipation,meili2010myosin,rape2011regulation,plotnikov2012force}. It also points to the possibility that a large part of the mechanical work due to actomyosin contractility is dissipated. Therefore, there is a need to discern the role of individual parameters in deciphering the mechanisms which regulate force generation. In this work, we have established the explicit role of myosin activity in generating rich dynamics within individual focal adhesion complexes, focusing our attention on a subset of the force regulation machinery involving the motor proteins and clutches, while ignoring the substrate elasticity. While applicable over a broad experimentally relevant parameter space, our model reproduces stick-slip type behaviour at lower active velocities and successfully demonstrates self-sustaining oscillations known to occur within FAs~\cite{plotnikov2012force,wu2017two}. Stochastic simulations of the system validate the existence of dynamical phases predicted by our model.

The coupled ordinary differential equations capture a coarse-grained picture of the biomechanical processes at play and act as a modular mechanism of traction force generation that can be combined to devise complex actomyosin networks which partake in durotaxis. Force fluctuations within FAs and concurrent oscillations in stress fibres (molecular motors) have been observed in experimental setups. Earlier theoretical models predicted spontaneous directed motions of motor proteins~\cite{julicher1995cooperative} and subsequently a stick-slip type dynamics with the motor-clutch paradigm~\cite{chan2008traction,bangasser2013master}. These models either assumed that the forces exerted by the stress fibres on FA are constant~\cite{sabass2010modeling,chan2008traction} or did not take the roles of myosin contractility and attachment-detachment dynamics into account~\cite{bangasser2013determinants}. Our model produces a rich array of dynamical phases for a wide range of biologically relevant parameters that are not directly accessible from these earlier models. 

Further,  an in-vitro experimental set-up by Plaçais et. al. ~\cite{placcais2009spontaneous} of a minimal actomyosin system was shown to give rise to spontaneous oscillations under elastic loading. In this set-up, a single actin filament was attached to a micron sized bead optically trapped while its other end interacts with myosin motors attached to a glass substrate. The system shows spontaneous oscillations for a set of parameters, such as the density of the motors and the stiffness of the optical trap.  This experiment directly corresponds to the special case of the permanently bound clutch which we have discussed in Fig. ~\ref{fig:trap}, the $n_c$ bound clutches giving rise to an effective elastic loading, as discussed in the experiment. 

\begin{figure}
	\centering
	\includegraphics[width=\textwidth]{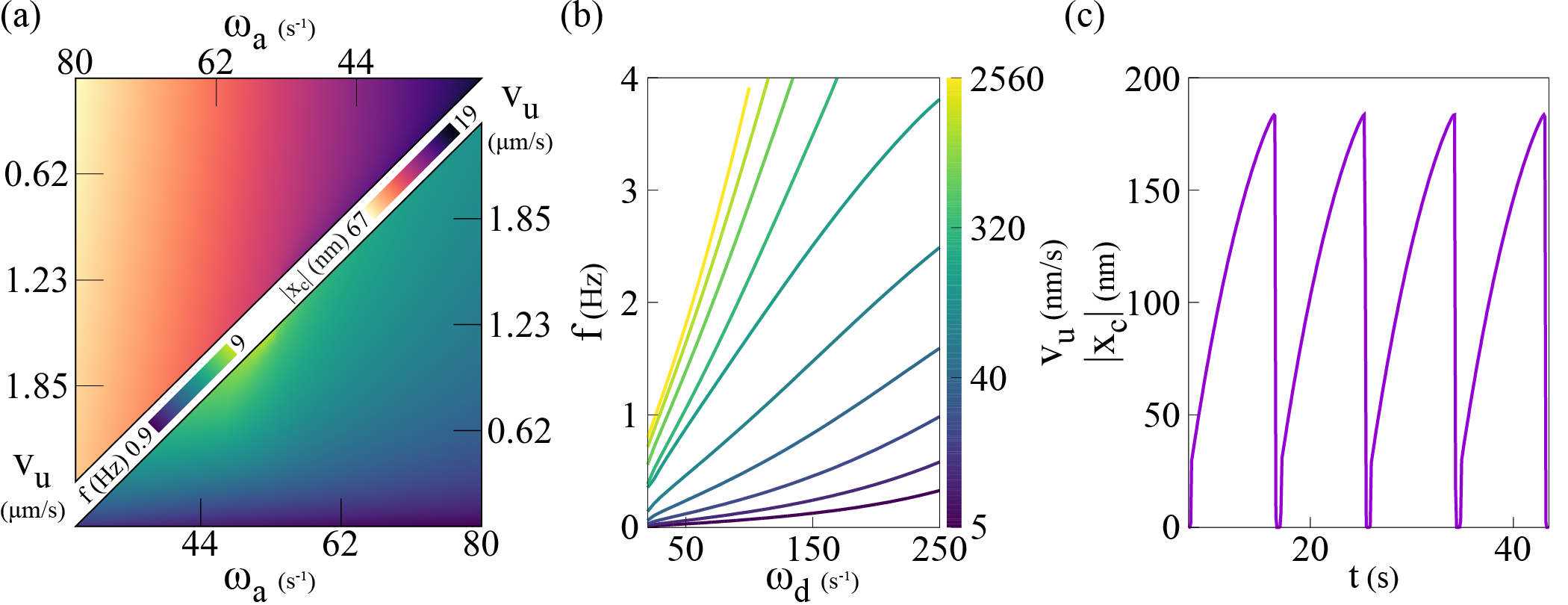}
	\caption{\textbf{Density plots characterising limit cycle oscillations:} With $N_c=10$, the upper and lower triangles in (a) delineate the amplitudes and frequencies $f$ of stable limit cycle oscillations in clutch extension $x_c$, respectively. The areas of these triangles are demarcated by the Hopf bifurcation boundary at their hypotenuse. \textbf{Dependence on motor detachment rate: }With $N_m=65$ and varying bare detachment rates, frequencies increase with rising $v_u$ within a range of $0\sim 4$ Hz, as seen in (b). The oscillations in clutch extension at $\omega_a=\omega_d$, and $v_u=0.025~\mu m/s$,  displays a frequency below $0.2$ Hz in (c), conforming with previous studies.}
	\label{fig:density}
\end{figure}

Our choice to focus on the role of myosin stems from experimental evidence of its role in the specific context of force fluctuations in individual focal adhesions and in regulating migration and mechanosensing~\cite{wu2017two,pasapera2015rac1,greenberg2016perspective}. Our attempt to recognise the key components that considerably influence the force fluctuations in FAs has resulted in a significant model simplification over that of Wu et al.~\cite{wu2017two} without sacrificing experimentally accessible parameters and results. The explicit role of myosin II in traction peak oscillations emerges as a natural consequence of the interaction between the motor and clutch sectors of the cellular migration machinery. In Fig.~\ref{fig:density}(a), we have plotted the frequencies and amplitude of limit cycle oscillations for a small number of clutches at intermediate MP velocity in physical units for the parameter values chosen in our study. With myosin II motor velocity varying in the range of $0.5-2~\mu$m/s, the oscillations in the average MP and clutch deformations vary in the range of $1-10$ Hz which is an order of magnitude higher than the typical oscillation frequencies in individual focal adhesions observed experimentally. However, the myosin detachment rates ($\omega_d$) that we considered in our study serves as an upper bound. The mechanochemical cycle in myosin II motors broadly consists of ATP hydrolysis followed by actin-binding, subsequent ADP release, and finally myosin detachment. Reduction in the ADP release rate or lower ATP concentrations significantly affects the bare detachment rate of myosin~\cite{wang2003kinetic}. 
In Fig.~\ref{fig:density}(b) we look at the change in the oscillation frequency for varying dissociation rates and for different values of the active velocity (all expressed in real units). We observe that the frequency for all values of the active velocity increases with increasing dissociation rate. There is a large range of dissociation rates and active velocities for which the frequency is in the range $0-1$ Hz, consistent with experimental and microscopic modelling results~\cite{wu2017two}.  In Fig. ~\ref{fig:density}(c), we provide an example of an oscillation in the clutch extension at experimentally observed values.

It is worthwhile to discuss the effect of introducing modifications in our model based on experimental observations. Both myosin and multiple components of the adhesion clutch (e.g., integrins, talin, and vinculin) have been shown to display catch-bond behaviour, i.e. an increase in bond lifetime with increasing load for a specific range of force.~\cite{sokurenko2008catch,thomas2008biophysics} For example, force applied to integrins has been shown to increase bond lifetimes by nearly an order of magnitude. Recent experiments have also suggested a direction-dependent catch-bond behaviour in the binding of vinculin~\cite{rothenberg2018vinculin}, which is a component of both cadherin and integrin based adhesion complexes, to actin filaments. Myosin II isoforms have also shown catch-bond behaviour, with the detachment rates of myosin varying accordingly.~\cite{stam2015isoforms,o2007myosin} These results emphasise the need to incorporate catch-bond in the detachment rates of both motor and clutch. Although we do not expect the mechanosensitivity of force generation to be affected by the catch bond kinetics, we do expect the oscillation frequencies to be sensitive to such behaviour. The presence of an external force or a substrate could change the timescales of attachment/detachment of motor and clutch proteins.

Substantially more critical is the effect of MP activity on the \textit{substrate deformation}, both by varying myosin velocity and attachment/detachment dynamics. Most theoretical studies which focus on force transmission in clutch models where substrate rigidity is tuned, show a biphasic relationship between substrate rigidity and force. This is understood in terms of a loading rate which is the speed at which forces in the clutches build when they are engaged and is directly controlled by the rigidity of the substrate~\cite{elosegui2018control}. 

Experimental observations suggest that force transmission is maximised for a specific value of rigidity or loading rate. Inhibiting myosin would lead to a decrease in the loading rate and therefore would require a higher rigidity of the substrate to reach the optimal value. Thus, although myosin inhibition leads to a reduction in myosin contractility, force transmission is enhanced for a range of rigidity.  This counterintuitive result has been shown experimentally~\cite{elosegui2016mechanical} using myosin inhibitor blebbistatin. Our model allows us to tune the myosin activity specifically via the myosin detachment rate and the myosin velocity, therefore providing a direct route to verify this counterintuitive result and predict an experimentally tunable parameter range to probe the mechanosensitivity of the molecular clutch.

\section*{Acknowledgements}

We acknowledge the use of computing facility at IISER Mohali. We thank Sudeshna Sinha and Debasish Chaudhuri for useful discussions. S.G. acknowledges QuantiXLie Centre of Excellence, a project co-financed by the Croatian Government and European Union through the European Regional Development Fund - the Competitiveness and Cohesion Operational Programme (Grant No. KK.01.1.1.01.0004).

\section*{Appendix A: Dimensionless Equations}\label{s:appxa}
Following the physical parameter values used in our model described in the Table \ref{tab:parameters}, we proceed to turn our dynamical equations dimensionless as prescribed in the main text. The characteristic scales for length, time, velocity, and force are calculated as $l_0 = 1.76392$ nm, $\omega_d^{-1}=0.00285714$ s, $v_0 = 617.373$ nm/s, and $f = 2.33854$ pN.

\begin{align}
  \frac{dn_m}{d\tau} &= \tilde{\omega}(N_m-n_m)-n_m\exp{\left(\frac{\tilde{\kappa}_m \tilde{y}}{\tilde{f}_d}\right)}\nonumber\\
  \frac{d\tilde{x}_c}{d\tau} &= -n_c\tilde{\kappa}_c\tilde{x}_c-n_m\tilde{\kappa}_m\tilde{y}\nonumber\\
  \frac{d\tilde{y}}{d\tau} &= \tilde{v}_u\left(1-\frac{\tilde{\kappa}_m\tilde{y}}{\tilde{f}_s}\right)+\frac{d\tilde{x}_c}{d\tau}\nonumber\\
  \frac{dn_c}{d\tau} &= \tilde{k}_{\text{on}}(N_c-n_c)-\tilde{k}_{\text{off}}n_c\exp{\left(\frac{-\tilde{\kappa}_c\tilde{x}_c}{\tilde{F}_b}\right)}
  \label{eq:dim_less_eqn}
\end{align}

\section*{Appendix B: Jacobian}\label{s:appxb}
The Jacobian matrix $(\mathcal{J})$ is computed to obtain the linearisation about the fixed points of the system which are calculated in the main text. The number of dynamical variables and concerning differential equations is 4, therefore the Jacobian matrix is of the order $4\times4$ and contains 16 elements as shown in Eq.~\eqref{eq:jack1}.
\begin{gather}
\frac{d}{d\tau}\begin{pmatrix}\tilde{x}_c\\\tilde{y}\\ n_m\\ n_c\end{pmatrix} =\mathcal{J}\begin{pmatrix}\tilde{x}_c\\\tilde{y}\\ n_m\\ n_c\end{pmatrix}=\begin{bmatrix}J_{11}&J_{12}&J_{13}&J_{14}\\J_{21}&J_{22}&J_{23}&J_{24}\\J_{31}&J_{32}&J_{33}&J_{34}\\J_{41}&J_{42}&J_{43}&J_{44}\end{bmatrix}\begin{pmatrix}\tilde{x}_c\\\tilde{y}\\n_m\\n_c\end{pmatrix}
\label{eq:jack1}
\end{gather}

The elements of the Jacobian matrix, $J_{ij}$, are explicitly calculated and the full matrix is depicted below,

\begin{align}
    \mathcal{J}=\begin{bmatrix}
    -\tilde{k}_{\text{on}}-\tilde{k}_{\text{off}}\exp{\left(\frac{n_m^0\tilde{f}_s}{n_c^0\tilde{F}_b}\right)} &\tilde{k}_{\text{off}}\frac{n_c^0\tilde{\kappa}_c}{\tilde{F}_b}\exp{\left(\frac{n_m^0\tilde{f}_s}{n_c^0\tilde{F}_b}\right)} & 0 & 0 \bigstrut \\
    \frac{\tilde{f}_s n_m^0}{n_c^0} & -n_c^0\tilde{\kappa}_c & -n_m^0\tilde\kappa_m & -\tilde{f}_s \bigstrut \\
    \frac{\tilde{f}_s n_m^0}{n_c^0} & -n_c^0\tilde{\kappa}_c & -\tilde{v}_u\frac{\tilde{\kappa}_m}{\tilde{f}_s}-n_m^0\tilde{\kappa}_m & -\tilde{f}_s\bigstrut \\
    0 & 0 & -n_m^0\frac{\tilde{\kappa}_m}{\tilde{f}_d}\exp{\left(\frac{\tilde{f}_s}{\tilde{f}_d}\right)} & -\tilde\omega-\exp{\left(\frac{\tilde{f}_s}{\tilde{f}_d}\right)} \bigstrut
\end{bmatrix}
\end{align}

The characteristic polynomial of the Jacobian has the form mentioned in Eq.~\eqref{eq:eigen_quartic},
\begin{align}
  P(\lambda) = \lambda^4+\mathcal{A}\lambda^3+\mathcal{B}\lambda^2+\mathcal{C}\lambda+\mathcal{D} = 0
  \label{charpol}
\end{align}

which is a fourth-order polynomial equation, where $\mathcal{A}$ is trace of matrix $\mathcal{J}$ or --Tr[$\mathcal{J}$], and $\mathcal{D}$ is determinant or Det[$\mathcal{J}$]. The coefficients are explicitly calculated as follows,

\begin{eqnarray}
  \mathcal{A} &=& \tilde{k}_{\text{on}} + \tilde{k}_{\text{off}}\exp{\left(\frac{n_m^0\tilde{f}_s}{n_c^0\tilde{F}_b}\right)} + n_c^0\tilde{\kappa}_c + \tilde v_u\frac{\tilde{\kappa}_m}{\tilde{f}_s} + n_m^0\tilde{\kappa}_m + \tilde\omega + \exp{\left(\frac{\tilde{f}_s}{\tilde{f}_d}\right)}\\
  \mathcal{B} &=& \frac{1}{\tilde{F}_b\tilde{f}_d\tilde{f}_s}\left[\exp\left\{\tilde{f}_s\left(\frac{1}{\tilde{f}_d}+\frac{n_m^0}{\tilde{F}_b n_c^0} \right) \right\}\tilde{F}_b\tilde{f}_d\tilde{f}_s\tilde{k}_{\text{off}} + \exp\left(\frac{\tilde{f}_s}{\tilde{f}_d}\right)\tilde{F}_b\left\{\tilde{f}_d\tilde{f}_s(\tilde{k}_{\text{on}}+n_c^0\tilde{\kappa}_c)+\tilde{f}_d\tilde{v}_u\tilde{\kappa}_m\right.\right.\nonumber\\
  &+&\left.(\tilde{f}_d-\tilde{f}_s)\tilde{f}_s n_m^0\tilde{\kappa}_m\right\}+\exp\left(\frac{\tilde{f}_s n_m^0}{\tilde{F}_b n_c^0}\right)\tilde{f}_d\tilde{k}_{\text{off}}\left\{\tilde{F}_b\tilde{v}_u\tilde\kappa_m-\tilde{f}_s^2 n_m^0\tilde\kappa_c+\tilde{F}_b\tilde{f}_s\left(n_c^0\tilde\kappa_c+n_m^0\tilde\kappa_m+\tilde\omega\right) \right\}\nonumber\\
  &+&\tilde{F}_b\tilde{f}_d\left\{\tilde{f}_s\tilde\omega(n_c^0\tilde\kappa_c+n_m^0\tilde\kappa_m)+\tilde{v}_u\tilde\kappa_m(\tilde{k}_{\text{on}}+n_c^0\tilde\kappa_c+\tilde\omega)+\tilde{f}_s\tilde{k}_{\text{on}}(n_c^0\tilde\kappa_c+n_m^0\tilde\kappa_m+\tilde\omega) \right\}\Bigg]\\
  \mathcal{C} &=& \frac{1}{\tilde{F}_b\tilde{f}_d\tilde{f}_s}\left[\exp\left\{\tilde{f}_s\left(\frac{1}{\tilde{f}_d}+\frac{n_m^0}{\tilde{F}_b n_c^0} \right) \right\}\tilde{k}_{\text{off}}\left[\tilde{F}_b\tilde\kappa_m\left\{(\tilde{f}_d-\tilde{f}_s)\tilde{f}_s n_m^0+\tilde{f}_d\tilde{v}_u\right\}-\tilde{f}_d\tilde{f}_s\tilde\kappa_c(\tilde{f}_s n_m^0-\tilde{F}_b n_c^0) \right]\right.\nonumber\\
  &+&\exp{\left(\frac{\tilde{f}_s}{\tilde{f}_d}\right)}\tilde{F}_b\left\{\tilde{f}_d\tilde{v}_u\tilde\kappa_m(\tilde{k}_{\text{on}}+n_c^0\tilde\kappa_c)+\tilde{f}_d\tilde{f}_s\tilde{k}_{\text{on}}(n_c^0\tilde\kappa_c+n_m^0\tilde\kappa_m)-\tilde{f}_s^2\tilde{k}_{\text{on}}n_m^0\tilde\kappa_m\right\}\nonumber\\
  &+&\tilde{F}_b\tilde{f}_d\left\{\tilde{k}_{\text{on}}n_c^0\tilde{v}_u\tilde\kappa_c\tilde\kappa_m+\tilde{v}_u\tilde\kappa_m\tilde\omega(\tilde{k}_{\text{on}}+n_c^0\tilde\kappa_c)+\tilde\omega\tilde{f}_s\tilde{k}_{\text{on}}(n_c^0\tilde\kappa_c+n_m^0\tilde\kappa_m) \right\}\nonumber\\
  &+&\left.\exp\left(\frac{\tilde{f}_s n_m^0}{\tilde{F}_b n_c^0}\right)\tilde{f}_d\tilde{k}_{\text{off}}\left\{\tilde{F}_b\tilde\kappa_m\tilde\omega\left(\tilde{v}_u+\tilde{f}_s n_m^0\right)+\tilde\kappa_c(\tilde{F}_b n_c^0-\tilde{f}_s n_m^0)(\tilde{v}_u\tilde\kappa_m+\tilde{f}_s\tilde\omega) \right\}\right]\\
  \mathcal{D} &=& \tilde v_u\frac{\tilde\kappa_c\tilde\kappa_m}{\tilde{f}_s\tilde{F}_b}\left\{\tilde\omega+\exp{\left(\frac{\tilde{f}_s}{\tilde{f}_d}\right)}\right\}\left[n_c^0\tilde{k}_{\text{on}}\tilde{F}_b+\tilde{k}_{\text{off}}\exp{\left(\frac{n_m^0 \tilde{f}_s}{n_c^0 \tilde{F}_b}\right)}\left(\tilde{F}_b n_c^0-\tilde{f}_s n_m^0\right)\right]
\end{eqnarray}

Nature and properties of the eigenvalues are dependent on the sign of the coefficients $\mathcal{A}$, $\mathcal{B}$, $\mathcal{C}$ and $\mathcal{D}$. We explore the algebra of polynomial equations to ascertain the features of the roots that they possess, which, in turn, provides us with the dynamical phases without explicitly solving the differential equations governing the system. In the following section, we shall detail a method to systematically determine the characteristics of algebraic roots of a real-valued polynomial equation.

\section*{Appendix C: Newton's rules for computing types and signs of roots}\label{s:appxc}
Newton formulated a set of rules that furnishes a lower bound for the cardinality of imaginary roots of a polynomial, in addition to the upper bound of positive roots, by taking into account the permanences and variations in an order of signs as procured from the polynomial.\\

Given a polynomial $P(x)$,
\begin{align}
  P(x) = \prescript{n}{}C_0 a_n x^n+\prescript{n}{}C_1 a_{n-1} x^{n-1}+\prescript{n}{}C_1 a_{n-1} x^{n-1}+\dots+\prescript{n}{}C_{n-1} a_{1} x+\prescript{n}{}C_0 a_{0} 
\end{align}

\textbf{Simple elements} are denoted as $a_n,\, a_{n-1},\, a_{n-2},\,\dots,\,a_1,a_0$. \textbf{Quadratic elements} are denoted as $Q_r$, where $Q_r$ is defined as follows,
\begin{align}
  \text{For} \quad P(x) = \sum^n_{i=0} p_{n-i} x^{n-i},\quad Q_r =& \frac{p_r^2}{{(\prescript{n}{}C_r)}^2}-\frac{p_{r+1}}{\prescript{n}{}C_{r+1}}\frac{p_{r-1}}{\prescript{n}{}C_{r-1}}\\
  =& \frac{1}{{(\prescript{n}{}C_r)}^2}\left[p_r^2-\frac{\prescript{n}{}C_r}{\prescript{n}{}C_{r+1}}\frac{\prescript{n}{}C_r}{\prescript{n}{}C_{r-1}}(p_{r+1})(p_{r-1}) \right]\nonumber\\
  =& \frac{1}{{(\prescript{n}{}C_r)}^2}\left[p_r^2-\frac{\frac{n!}{r!(n-r)!}}{\frac{n!}{(r+1)!(n-r-1)!}}\frac{\frac{n!}{r!(n-r)!}}{\frac{n!}{(r-1)!(n-r+1)!}} (p_{r+1})(p_{r-1}) \right]\nonumber\\
  \text{Finally,}\quad Q_r =&\frac{1}{{(\prescript{n}{}C_r)}^2}\left[p_r^2-\frac{r+1}{n-r}\frac{n-r+1}{r}(p_{r+1})(p_{r-1})\right]
\end{align}

\begin{theorem}[Newton's Incomplete Rule]
Supposing that the quadratic elements for a polynomial $P(x)$ are all non-zero, the number of variations of signs in the sequence $Q_n,Q_{n-1},\ldots ,Q_0$ provides a lower bound for the number of imaginary roots of $P(x)$.
\end{theorem}

To obtain Newton's complete rule, one has to look at the sequences of both simple and quadratic elements,
\begin{center}
  \begin{tabular}{cccccc}
  $a_n$ & $a_{n-1}$ & $a_{n-2}$ &\dots &$a_1$ & $a_0$\\
  $Q_n$ & $Q_{n-1}$ & $Q_{n-2}$ &\dots &$Q_1$ & $Q_0$ 
  \end{tabular}
\end{center}

By concentrating on associated pairs i.e.,
\begin{center}
  \begin{tabular}{cccc}
  \dots & $a_{r+1}$ & $a_r$ & \dots\\
  \dots & $Q_{r+1}$ & $Q_r$ & \dots 
  \end{tabular}
\end{center}

We are to look for possibilities of sign changes in the aforementioned pair by denoting them by their permanence, i.e. no changes in sign and variance, i.e. changes in sign in the following manner: a lowercase \textbf{v} denotes variance in sign of upper element of the pairs, an uppercase \textbf{V} denotes variance in the sign of lower element of the pairs, a lowercase \textbf{p} denotes permanence of sign of upper element of the pairs and an uppercase \textbf{P} denotes permanence of sign of lower element of the pairs. By instating this schema, we obtain four possible ways the signs can change in a pair --- vV, vP, pV and pP.

\begin{theorem}[Newton's Complete Rule]
Supposing a non zero simple and quadratic elements of $P(x)$, then the total number of double permanences, written as $\sum pP$ is an upper bound of number of negative roots and total number of variance-permanences, written as $\sum vP$ is the upper bound of positive roots.
\end{theorem}
\begin{corollary}
  The total number of real roots are the sum of double permanences and variance-permanences.
\end{corollary}

Therefore, the total number of real roots is equal to the total number of permanences in quadratic elements i.e. $\sum P$. This is an upper bound of the real roots. Thus $n-\sum P = \sum V$ is the lower bound of number of complex roots. We may now proceed with using these rules to obtain the bounds on types of roots for a quartic polynomial with real coefficients that appears as a characteristic polynomial for our system.\\

A quartic polynomial $P_4(x)$ has the following form,
\begin{align}
    P_4(x) &= \prescript{4}{}C_0 a_4 x^4 + \prescript{4}{}C_1 a_3 x^3 + \prescript{4}{}C_2 a_2 x^2 + \prescript{4}{}C_3 a_1 x + \prescript{4}{}C_4 a_0\nonumber \\
    &= a_4 x^4 + 4a_3 x^3 + 6 a_2 x^2 + 4 a_1 x + a_0
\end{align}

Comparing it with the quartic polynomial of the form $x^4 + \mathcal{A}x^3 + \mathcal{B}x^2 + \mathcal{C}x+ \mathcal{D}$, as used in the main text, the simple elements are calculated to be $a_4 = 1, a_3 = \mathcal{A}/4, a_2 = \mathcal{B}/6, a_1 = \mathcal{C}/4$ and finally $a_0 = \mathcal{D}$. Similarly, the quadratic elements are $Q_4 = 1, Q_3 = \mathcal{A}^2/16- \mathcal{B}/6, Q_2 = \mathcal{B}^2/36 - \mathcal{AC}/16, Q_1 = \mathcal{C}^2/16 - \mathcal{BD}/6$ and $Q_0 = \mathcal{D}^2$.\\

It is possible to numerically show that any quartic polynomial will have at most 14 different combinations of roots. Our system has two constraints on the characteristic polynomial due to the fact that two of the coefficients, $\mathcal{A}$ and $\mathcal{D}$ are entirely positive inside the relevant parametric space, thus leaving only 4 possible combinations of signs for $\mathcal{B}$ and $\mathcal{C}$, as we shall observe. The coefficients of the characteristic polynomial have the following limits in the parametric space,

\begin{table}
  \centering
  \begin{tabular}{rcr}
  \toprule
  Lower Limit & Coefficient & Upper Limit\bigstrut\\
  \midrule
  8.35772 & $\mathcal{A}$ & 10.2638 \bigstrut \\
  -2.53967 & $\mathcal{B}$ & 13.2611 \bigstrut \\
  -0.0142965 & $\mathcal{C}$ & 4.15717 \bigstrut \\
  0 & $\mathcal{D}$ & 0.0129296 \\
  \bottomrule
  \end{tabular}
  \caption{\textsf{The limits on the values of coefficients $\mathcal{A,B,C \text{ \& } D}$}}
  \label{tab:tables2}
\end{table}

We proceed with finding the bounds on cardinality of different types of roots for our system by calculating the simple and quadratic elements as described earlier with different combinations of coefficients under the bounds laid down in Table \ref{tab:tables2}.

\begin{enumerate}[{CASE} I]
    \item \textsf{\large{Both $\mathcal{B}$ and $\mathcal{C}$ are positive}}\\\\
    The simple elements do not have a change in sign which prohibits roots with positive $\mathbb{R}$ part. $\Sigma vP$ being zero throughout confirms this.
    \begin{center}
        \begin{tabularx}{\textwidth}{>{\columncolor{darkgray}}c|ccccc|l}
          {\color{white}a} & + & + & + & + & + & \multirow{2}{25em}{$\Sigma pP$ is 4, either 2 (--) $\mathbb{R}$ roots \& 2 $\mathbb{C}$ roots with (--) $\mathbb{R}$ part, or 4 (--) $\mathbb{R}$ roots}\\
          {\color{white}Q} & + & + & + & + & + &
        \end{tabularx}
    \end{center}
    \begin{center}
        \begin{tabularx}{\textwidth}{>{\columncolor{darkgray}}c|ccccc|l}
          {\color{white}a} &  + & + & + & + & + & \multirow{2}{25em}{$\Sigma pP$ is 2, 2 (--) $\mathbb{R}$ roots \& 2 $\mathbb{C}$ roots with (--) $\mathbb{R}$ part}\\
          {\color{white}Q} & + & + & + & -- & + &
        \end{tabularx}
    \end{center}
    \begin{center}
        \begin{tabularx}{\textwidth}{>{\columncolor{darkgray}}c|ccccc|l}
          {\color{white}a} &  + & + & + & + & + & \multirow{2}{25em}{$\Sigma pP$ is 2, 2 (--) $\mathbb{R}$ roots \& 2 $\mathbb{C}$ roots with (--) $\mathbb{R}$ part}\\
          {\color{white}Q} & + & + & -- & + & + &
        \end{tabularx}
    \end{center}
    \begin{center}
        \begin{tabularx}{\textwidth}{>{\columncolor{darkgray}}c|ccccc|l}
          {\color{white}a} &  + & + & + & + & + & \multirow{2}{25em}{$\Sigma pP$ is 2, 2 (--) $\mathbb{R}$ roots \& 2 $\mathbb{C}$ roots with (--) $\mathbb{R}$ part}\\
          {\color{white}Q} & + & + & -- & -- & + &
        \end{tabularx}
    \end{center}

    \item \textsf{\large{$\mathcal{B}$ is positive but $\mathcal{C}$ is negative}}
    \begin{center}
        \begin{tabularx}{\textwidth}{>{\columncolor{darkgray}}c|ccccc|l}
          {\color{white}a} &  + & + & + & -- & + & \multirow{2}{25em}{$\Sigma pP = \Sigma vP$ = 2, maximum 2 (+) and 2 (--) $\mathbb{R}$ roots}\\
          {\color{white}Q} & + & + & + & + & + &
        \end{tabularx}
    \end{center}
    \begin{center}
        \begin{tabularx}{\textwidth}{>{\columncolor{darkgray}}c|ccccc|l}
          {\color{white}a} &  + & + & + & -- & + & \multirow{2}{25em}{$\Sigma pP$ is 2, and $\Sigma vP$ is 0, i.e. maximum 2 (--) $\mathbb{R}$ roots but no (+) $\mathbb{R}$ roots}\\
          {\color{white}Q} & + & + & + & -- & + &
        \end{tabularx}
    \end{center}

    \item \textsf{\large{$\mathcal{B}$ is negative but $\mathcal{C}$ is positive}}
    \begin{center}
        \begin{tabularx}{\textwidth}{>{\columncolor{darkgray}}c|ccccc|l}
          {\color{white}a} &  + & + & -- & + & + & \multirow{2}{25em}{$\Sigma pP = \Sigma vP$ = 2, maximum 2 (+) and 2 (--) $\mathbb{R}$ roots}\\
          {\color{white}Q} & + & + & + & + & + &
        \end{tabularx}
    \end{center}
    \begin{center}
        \begin{tabularx}{\textwidth}{>{\columncolor{darkgray}}c|ccccc|l}
          {\color{white}a} &  + & + & -- & + & + & \multirow{2}{25em}{$\Sigma pP$ is 2, and $\Sigma vP$ is 0, i.e. maximum 2 (--) $\mathbb{R}$ roots but no (+) $\mathbb{R}$ roots}\\
          {\color{white}Q} & + & + & -- & + & + &
        \end{tabularx}
    \end{center}

    \item \textsf{\large{Both $\mathcal{B}$ and $\mathcal{C}$ are negative}}
    \begin{center}
        \begin{tabularx}{\textwidth}{>{\columncolor{darkgray}}c|ccccc|l}
          {\color{white}a} &  + & + & -- & -- & + & \multirow{2}{25em}{$\Sigma pP = \Sigma vP$ = 2, maximum 2 (+) and 2 (--) $\mathbb{R}$ roots}\\
          {\color{white}Q} & + & + & + & + & + &
        \end{tabularx}
    \end{center}    

\end{enumerate}

We proceed to collate various possible combinations of roots, as predicted by Newton's rules of signs, in Table \ref{tab:tables3}.\\

From Table \ref{tab:tables3}, we can conclude, with $\lambda_j$s, where $j=1,...,4$, denoting four eigenvalues, miscellany of positive and negative $\mathcal{B}$, and $\mathcal{C}$ lead to the following combination of eigenvalues: (i) $\lambda_{1,2,3,4}$ all are real negative, (ii) $\lambda_{1,2}$ real negative and $\lambda_{3,4}$ real positive, (iii) $\lambda_{1,2}$ real negative and $\lambda_{3,4} = -\alpha\pm i\beta$, and (iv) $\lambda_{1,2}$ real negative and $\lambda_{3,4} = \alpha\pm i\beta$. $\alpha$ and $\beta$ are real positive numbers. Case (i) corresponds to linearly stable (s) phase where a perturbation decays exponentially with time and the system returns to its fixed point. Case (ii) is characterised by exponentially growing perturbations in time and called unstable (u) phases. Instability in our system is established when all the clutches are detached from the actin filament and it is is freely pulled by the molecular motors. Stable spiral (ss) or oscillation decaying with time is the characteristic property of case (iii), which reaches stable (s) phase at long time scale. Growing oscillation in time is a hallmark of unstable spiral (us) that originates from the presence of positive real part of the complex eigenvalues as indicated in (iv). Going beyond the ambit of linear stability and numerically solving the coupled non-linear equations, presents the unstable spiral phase as a precursor of stable oscillation in the system, as shown in Fig.~\ref{fig:phased}

\begin{table}
    \centering
  
    \begin{tabularx}{\textwidth}{ZZZZZZ}
      \toprule
      \multicolumn{2}{c}{Signs of Coefficients} & \multicolumn{4}{c}{Types of Roots} \bigstrut \\
      \midrule
      $\mathcal{B}$ & $\mathcal{C}$ & $\lambda_1$ & $\lambda_2$ & $\lambda_3$ & $\lambda_4$ \\
      \midrule
      \multirow{2}{*}{+} & \multirow{2}{*}{+}& $-\mathbb{R}$ & $-\mathbb{R}$ & $-\mathbb{R}$ & $-\mathbb{R}$\\
      & & $-\mathbb{R}$ & $-\mathbb{R}$ & $-\mathfrak{R}(\mathbb{C})$ & $-\mathfrak{R}(\mathbb{C})$\\
      \midrule
      \multirow{2}{*}{+} & \multirow{2}{*}{$-$}& $-\mathbb{R}$ & $-\mathbb{R}$ & $+\mathbb{R}$ & $+\mathbb{R}$\\
      & & $-\mathbb{R}$ & $-\mathbb{R}$ & $+\mathfrak{R}(\mathbb{C})$ & $+\mathfrak{R}(\mathbb{C})$\\
      \midrule
      \multirow{2}{*}{$-$} & \multirow{2}{*}{+}& $-\mathbb{R}$ & $-\mathbb{R}$ & $+\mathbb{R}$ & $+\mathbb{R}$\\
      & & $-\mathbb{R}$ & $-\mathbb{R}$ & $+\mathfrak{R}(\mathbb{C})$ & $+\mathfrak{R}(\mathbb{C})$\\
      \midrule
      \multirow{2}{*}{$-$} & \multirow{2}{*}{$-$}& $-\mathbb{R}$ & $-\mathbb{R}$ & $+\mathbb{R}$ & $+\mathbb{R}$\\
      & & $-\mathbb{R}$ & $-\mathbb{R}$ & $+\mathfrak{R}(\mathbb{C})$ & $+\mathfrak{R}(\mathbb{C})$\\
      \bottomrule
  
    \end{tabularx}
    \caption{\textsf{Possible roots ($\lambda_i$) resulting from various combination of signs of $\mathcal{B}$ and $\mathcal{C}$. $\mathbb{R}$ denotes real roots and $\mathfrak{R}(\mathbb{C})$ refers to real parts of complex roots.}}
    \label{tab:tables3}
\end{table}

\section*{Appendix D: Coefficients for the case where the clutches are always bound to the filament}\label{s:appxd}
The coefficients ${\cal A}^{\prime}, {\cal B}^{\prime}$ and ${\cal C}^{\prime}$ are given as follows:
\begin{align}
  \mathcal{A}' &= \tilde\omega + \exp{\left(\frac{\tilde{f}_s}{\tilde{f}_d}\right)} + \frac{1}{\epsilon^2}\left(N_c\tilde{\kappa}_c + n_m^0\tilde{\kappa}_m\right)+\frac{\tilde{v}_u\tilde{\kappa}_m}{\tilde{f}_s}\\ 
  \mathcal{B}' &= \frac{\tilde{v}_u\tilde{\kappa}_m N_c\tilde{\kappa}_c}{\tilde{f}_s\epsilon^2} + \left[\tilde\omega + \exp{\left(\frac{\tilde{f}_s}{\tilde{f}_d}\right)}\right]\left[\frac{\tilde{v}_u\tilde{\kappa}_m}{\tilde{f}_s} + \frac{N_c\tilde{\kappa}_c}{\epsilon^2} + \frac{n_m^0\tilde\kappa_m}{\epsilon^2}\right] - \frac{\tilde{f}_s n_m^0 \tilde\kappa_m}{\epsilon^2\tilde{f}_d}\exp{\left(\frac{\tilde{f}_s}{\tilde{f}_d} \right)}\\
  \mathcal{C}' &= \frac{\tilde{v}_u\tilde{\kappa}_m N_c\tilde{\kappa}_c}{\tilde{f}_s\epsilon^2}\left[\tilde\omega + \exp{\left(\frac{\tilde{f}_s}{\tilde{f}_d}\right)}\right]
\end{align}

{\footnotesize \sffamily \bibliography{ms}}
\bibliographystyle{science}
\end{document}